\documentclass[preprint,12pt]{elsarticle}
\usepackage{amssymb}
\newcommand{\beqn}{\begin{eqnarray}}
\newcommand{\eeqn}{\end{eqnarray}}
\newcommand{\eq}[1]{(\ref{#1})}

\newcommand{\tr}{ {\rm Tr} \, }

\newcommand{\Z}{{\mathbb Z}}

\newcommand{\lr}[1]{ \left( #1 \right) }
\newcommand{\lrs}[1]{ \left[ #1 \right] }

\newcommand{\vev}[1]{ \langle \, #1 \, \rangle }
\journal{Physics Letters B}
\begin{document}

\begin{frontmatter}
\title{Electromagnetic superconductivity of vacuum induced by strong magnetic field: numerical evidence in lattice gauge theory}

\author[IHEP]{V. V. Braguta}
\author[ITEP,JINR]{P. V. Buividovich}
\author[CNRS,GENT]{M. N. Chernodub\fnref{fn1}}
\author[ITEP,MIPT]{A.~Yu.~Kotov}
\author[ITEP,MIPT]{M. I. Polikarpov}

\address[IHEP]{IHEP, Protvino, Moscow region, 142284 Russia}
\address[ITEP]{ITEP, B. Cheremushkinskaya str. 25, Moscow, 117218 Russia}
\address[JINR]{JINR, Joliot-Curie str. 6, Dubna, Moscow region, 141980 Russia}
\address[CNRS]{CNRS, Laboratoire de Math\'ematiques et Physique Th\'eorique, Universit\'e Fran\c{c}ois-Rabelais Tours, Parc de Grandmont, 37200 Tours, France}
\address[GENT]{Department of Physics and Astronomy, University of Gent, Krijgslaan 281, S9, B-9000 Gent, Belgium}
\address[MIPT]{MIPT, Institutskii per. 9, Dolgoprudny, Moscow Region, 141700 Russia}

\fntext[fn1]{On leave from ITEP, Moscow, Russia.}

\begin{abstract}
Using numerical simulations of quenched SU(2) gauge theory we demonstrate that an external magnetic field leads to spontaneous generation of quark condensates with quantum numbers of electrically charged $\rho$ mesons if the strength of the magnetic field exceeds the critical value $e B_c = 0.927(77)\, {\mbox{GeV}}^2$ or $B_c =(1.56 \pm 0.13)\cdot 10^{16}\, {\mbox{Tesla}}$. The condensation of the charged $\rho$ mesons in strong magnetic field is a key feature of the magnetic-field-induced electromagnetic superconductivity of the vacuum.
\end{abstract}


\end{frontmatter}

Effects caused by very strong magnetic fields attract increasing interest motivated by the fact that the hadron-scale strong magnetic fields may emerge in the heavy-ion collisions at the Relativistic Heavy-Ion Collider at Brookhaven National Laboratory and at the Large Hadron Collider at CERN~\cite{ref:Skokov}. Such fields may, presumably, have arisen in the early Universe~\cite{ref:Universe}. 

The strong magnetic field causes exotic effects in hot quark-gluon matter, the well-known example is the chiral magnetic effect~\cite{ref:CME}. In the absence of matter the  magnetic field background leads also to unusual effects like the magnetic catalysis~\cite{ref:catalysis}, shift of finite-temperature transitions in QCD~\cite{ref:transition} and anisotropic conductivity~\cite{ref:conductivity}.

It was recently suggested that a sufficiently strong external magnetic field should turn the vacuum into an electromagnetic superconductor~\cite{ref:PRD,ref:PRL}. The superconductivity emerges due to spontaneous condensation of electrically charged vector particles, $\rho^\pm$~mesons,  if the magnetic field exceeds the critical strength
\beqn
e B_c = m^2_\rho \approx 0.6 \, \mathrm{GeV}^{2},
\quad
B_c = m^2_\rho / e \approx 10^{16} \, \mathrm{T}, \quad
\label{eq:Bc:theor}
\eeqn
where $m_\rho = 775.5\,{\mathrm{MeV}}$ is the mass of the $\rho$ meson and $e$ is the elementary electric charge. In terms of the up-quark ($u$) and down-quark ($d$) fields the suggested $\rho$-meson condensate should have the following form~\cite{ref:PRL}:
\beqn
\langle \bar u \gamma_1 d\rangle = - i \langle \bar u \gamma_2 d\rangle = \rho\,, \qquad
\langle \bar u \gamma_3 d\rangle = \langle \bar u \gamma_0 d\rangle = 0\,, \quad
\label{eq:ud:cond}
\eeqn
where $\rho = \rho(x^\perp)$ is a certain complex-valued periodic function of the coordinates $x^\perp = (x^1,x^2)$ of a plane which is perpendicular to the magnetic field $\vec B = (0,0,B)$.

The superconducting vacuum should have many unusual features. Firstly, no matter is required to create the superconductor so that the electromagnetic superconductivity appears literally "from nothing". Secondly, the superconductivity is anisotropic so that the vacuum acts as a superconductor along the magnetic-field axis only. Thirdly, the superconductivity is inhomogeneous because the $\rho$-meson condensate is not uniform in the $\vec B$-transverse directions  due to the presence a new type of topological defects, the $\rho$ vortices. Fourthly, the net electric charge of the superconducting vacuum is zero despite of the presence of the charged condensates~\eq{eq:ud:cond}~\cite{ref:PRD,ref:PRL}. 

The spontaneous generation of the $\rho$-meson condensate~\eq{eq:ud:cond} -- which plays a role of the Cooper pair condensate in the conventional superconductivity -- is the key feature of the vacuum superconductor mechanism~\cite{ref:PRD,ref:PRL}.

The appearance of the $\rho$-meson condensate was found analytically in a phenomenological model based on the vector meson dominance (VMD)~\cite{ref:PRD}, in the Nambu--Jona-Lasinio (NJL) model~\cite{ref:PRL}, and in holographical approaches based on gauge/gravity duality~\cite{ref:holography}. We use numerical simulations of the lattice gauge theory to demonstrate that strong magnetic field indeed leads to emergence of the superconducting condensate of the charged $\rho$ mesons.

In QCD the charged $\rho$ meson field is identified with $\rho_\mu = \bar u \gamma_\mu d$.
The condensation pattern~\eq{eq:ud:cond} corresponds to the condensate of the $\rho$ mesons with the spins aligned along the axis of the magnetic field, with the $s_z=+1$ projection of the spin onto the $z$ axis. It is convenient to introduce two combinations of the negatively--charged $\rho$-meson fields\footnote{One can equivalently work with the positively--charged $\rho$--meson fields, $\bar d (\gamma_1 \pm i \gamma_2) u$. The magnitudes of the positive and negative condensate are equivalent because the vacuum state is electrically neutral. Our results on vacuum condensation are the same for positive and negatively charged operators.},
\beqn
\rho_\pm = \frac{1}{2} (\rho_1 \pm i \rho_2) \equiv  \frac{1}{2} \bar u (\gamma_1 \pm i \gamma_2) d\,,
\label{eq:rho:sz}
\eeqn
which correspond to the spin projections $s_z=\pm 1$, respectively. Indeed, according to the simplified arguments of Ref.~\cite{ref:PRD}, the invariant masses $M_{n,s_z}$ of the $\rho$ mesons states in the magnetic field $B$ should behave as follows,
\beqn
M^2_{n,s_z} = m^2_\rho + (1+2 n - 2 s_z) |eB|\,,
\label{eq:M}
\eeqn
with the nonnegative integer $n$ and the spin projection onto the $z$ axis $s_z = -1, 0, +1$. The ground state is identified with the quantum numbers $n=0$ and $s_z = +1$, and the charged $\rho$ mesons should get condensed, $M_{0,+1}^2 < 0$, if the  magnetic field exceeds the critical value~\eq{eq:Bc:theor}.

The simplest way to check numerically the possible appearance of the superconducting condensate~\eq{eq:ud:cond} is to calculate the equal-time correlation function along the direction of the magnetic field, 
\beqn
G_\pm(z) = \langle \rho^\dagger_\pm(0) \rho_\pm(z) \rangle\,,
\label{eq:G:pm}
\eeqn
where the separation in the transverse coordinates of the two probes is set to zero, $x^\perp = (0,0)$. The long-distance behavior of the $s_z = +1$ correlation function~\eq{eq:G:pm} should expose the expected emergence of the condensate~\eq{eq:ud:cond} due to the factorization property:
\beqn
\lim\nolimits_{|z| \to \infty} G_+ (z) = |\langle \rho \rangle|^2\,,
\label{eq:factorization}
\eeqn
while the $\rho$ mesons with the opposite orientation of the spins, $s_z=-1$, should not be condensed\footnote{The condensed component is identified by the positiveness of the product $e B s_z$. If the magnetic field is reversed and becomes negative, $e B < 0$, then the condensed component corresponds to $s_z = -1$  while the component with $s_z = +1$ is not condensed.}:
\beqn
\lim\nolimits_{|z| \to \infty} G_- (z) = 0\,.
\label{eq:factorization:minus}
\eeqn

We calculate the correlation functions~\eq{eq:G:pm} numerically, using lattice Monte-Carlo simulations of quenched $SU(2)$ lattice gauge theory following numerical setup
of Ref.~\cite{ref:conductivity}. The quark fields are introduced by the overlap lattice Dirac operator $\mathcal{D}$ with exact chiral symmetry \cite{Neuberger:98:1}. 
The correlation function~\eq{eq:G:pm} is a linear combination of the current-current correlators in the vector meson channel. The vector correlator is represented in terms of Dirac propagators in fixed background of Abelian and non-Abelian gauge fields and is then averaged over an equilibrium ensemble of non-Abelian gauge fields $A_\mu$:
\begin{eqnarray}
\label{four_fermion_vev}
& & \vev{\bar{u}\lr{x} \gamma_{\mu} d\lr{x} \, \bar{d}\lr{y} \gamma_{\nu} u\lr{y}} = 
\left(\int D A_{\mu}\,e^{-S_{YM}\lrs{A_{\mu}}}\right)^{-1}
\nonumber \\ & & \cdot
\int D A_{\mu}\,e^{-S_{YM}\lrs{A_{\mu}}}\, \tr\lr{\frac{1}{\mathcal{D}_u + m} \, \gamma_{\mu} \, \frac{1}{\mathcal{D}_d + m} \, \gamma_{\nu}},
\qquad
\end{eqnarray}
where $S_{YM}\lrs{A_{\mu}}$ is the lattice action for gluons $A_{\mu}$.

A uniform time-independent magnetic field $B$ is introduced into the Dirac operator ${\mathcal{D}}_f$ for the flavor $f=u,d$ in a standard way by substituting the $su(2)$-valued vector potential $A_{\mu}$ with the $u(2)$-valued potential $A_{\mu \, ij} \rightarrow A_{\mu \, ij} +  \delta_{ij} q_f\: F_{\mu\nu}\: x_{\nu}/2$,
where $q_f$ is the electric charge of the corresponding quark, $q_u = +  2 e/3$, $q_d = - e/3$, and $i,j$ are color indices. We also introduce an additional twist for fermions in order to account for periodic boundary conditions in spatial directions~\cite{Wiese:08:1,ref:CME:lattice}. For technical reasons the bare quark mass $m_0$ is fixed at a small value $a m_0 = 0.01$, where $a$ is the lattice spacing. The vector correlation functions depend very weakly on the bare quark mass if they are calculated with the help of the overlap Dirac operator~\cite{Babich:05:1}.

Our numerical approach is done in two complimentary ways. Firstly, we study in details the superconducting condensate for eleven values of the magnetic field $B$ at a relatively small symmetric $14^4$ lattice at a single lattice gauge coupling $\beta = 3.281$. These parameters correspond to the physical volume of the lattice is $L^{4} = (1.44\, {\mathrm{fm}})^4$ and the lattice spacing $a=0.103\, {\mathrm{fm}}$~\cite{Bornyakov:2005iy}. Then we use an heuristic fitting method to find the superconducting condensate without taking a long-range limit~\eq{eq:factorization} because the factorization~\eq{eq:factorization} does not work well in too small volume. Secondly, we consider a set of lattices of various physical volumes $L^{4}$ and four values of magnetic field strengths $B$, and then utilize a conventional fitting procedure to extract the condensate $\eta = \eta(L)$. The extrapolation to the infinite volume, $L \equiv l a \to\infty$, shows that these two methods give the same results. In both approaches the ultraviolet lattice artifacts are reduced with the help of the tadpole-improved Symanzik gluon action~\cite{ref:improved}.

In our first approach we use 30 configurations of the gluon gauge field for each value of the background magnetic field. The periodicity of the lattice leads to the quantization ($k \in \Z$) of the magnetic field,
\beqn
B = k \, B_{\mathrm{min}}\,, \qquad 
e B_{\mathrm{min}} = \frac{3 \cdot 2 \pi}{L^2} = 0.354\, {\mathrm{GeV}}^2\,,
\label{eq:B:quant}
\label{eq:eB:min}
\eeqn
because of the requirement  
\beqn
\int d^2 x^\perp \, q_f B \in \Z \quad {\mbox{for}} \quad f=u,d\,.
\eeqn 
In Eq.~\eq{eq:B:quant} the integer $k = 0, 1, \dots, L_s^2/2$ determines the number of elementary magnetic fluxes which pass through the boundary of the lattice in the $(x^1,x^2)$ plane. 

The maximal possible value of the fluxes $k = l^2/2 = 98$ corresponds to an extremely large magnetic field with the magnetic length $L_B \sim (e B)^{-1/2}$ being of the order of the lattice spacing, $L_B \sim a$. In order to avoid associated ultraviolet artifacts, in our simulations we limit the maximal value of the fluxes by $k_{\mathrm{max}} = 10 \ll l^2/2$, so that our maximal magnetic field, $e B_{\mathrm{max}} = 3.54\, {\mathrm{GeV}}^2$ is much larger than the expected critical magnetic field~\eq{eq:Bc:theor}.

\begin{figure}[!thb]
\begin{center}
\includegraphics[scale=0.9,clip=false]{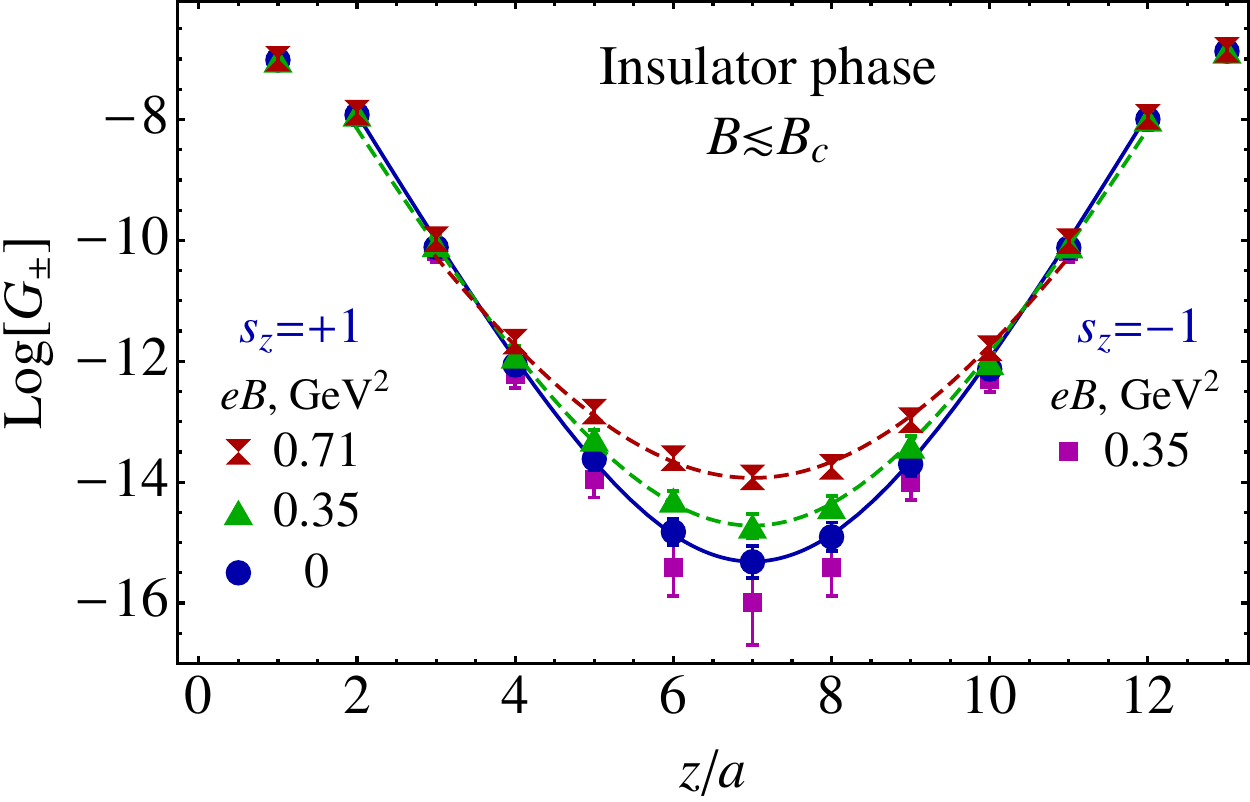}
\end{center}
\caption{The $G_+$ correlator ($s_z {=} + 1$) at $eB {=} 0,$ $0.35,$ $0.71$\,GeV${}^2$  and
the $G_-$ correlator ($s_z{=} - 1$) at $eB{=} 0.35$\,GeV${}^2$ in the insulator (weak magnetic field) phase
(in lattice units). The lines illustrate the best fits by the heuristic function~\eq{eq:G:fit:weak}.}
\label{fig:weak}
\end{figure}
In Fig.~\ref{fig:weak} we show correlator~\eq{eq:G:pm} for a few relatively small values of the magnetic field. As we have anticipated, the $s_z {=}\pm 1$ correlators split in the external magnetic field [in Fig.~\ref{fig:weak} we show both spin orientations for $B = B_{\mathrm{min}}$]. The observed splitting reflects the change in the relevant lowest energies \eq{eq:M}, $M^2_\pm = m^2_\rho \mp |eB|$, so that the expected hierarchy of the masses, $M_+ {<} m_\rho {<} M_-$, is encoded in the slopes of the correlators $$G_\pm(z) \sim e^{- M_\pm |z|} + \cdots.$$ The splitting of the $s_z=\pm 1$ masses was also found in $SU(3)$ lattice gauge theory at weak magnetic fields~\cite{ref:lee}. We have checked that the no-condensation property~\eq{eq:factorization:minus} for the $s_z{=}-1$ correlator $G_-$ is valid for all studied values of the magnetic field.

In the weak field region, $B < B_{c}$, the long distance be\-havior of the correlator $G_+$ is expected to be proportional to the function $e^{-\mu |z|}$, or $\cosh [(|z| - L/2) \mu ]$ in a finite volume with periodic boundary conditions. Here $\mu$ is a massive parameter and $L = l a$ is the physical lattice size.

We have found, however, that our numerical data for the correlators~\eq{eq:G:pm} are consistent with the cosh-like behavior only in a very narrow interval of the coordinate $z$. Therefore, in at  $B < B_{c}$ we used a heuristic fit function 
\beqn
G^{\mathrm{fit, weak}}_+ = C_{\mathrm{weak}} e^{-\mu \gamma L/2}  \cosh^\gamma [\mu (|z| - L/2)]\,,
\label{eq:G:fit:weak}
\eeqn
which reduces to the exponential (``no-condensate'') behavior in the thermodynamic limit $L \to \infty$. In Eq.~\eq{eq:G:fit:weak} $\mu> 0$, $\gamma> 0$ and $C_{\mathrm{weak}}>0$ are the fitting parameters. 

\begin{table}
\begin{tabular}{c | c l l || l l l l l l}
$eB$, GeV${}^{2}$    & 0   & 0.35 & 0.71 & 1.06 & 1.42 & 1.77 & 2.12 & 2.83 & 3.54\\
\hline
weak                            &0.30& 0.46 & 1.8 & 4.5 & 5.8   &  5.8    &  5.8  &  5.4  & 5.0 \\
strong                          &    -   &   -     &  -      & 2.1  & 1.9 &1.1 & 0.23 & 0.25 & 0.37 
\end{tabular}
\caption{Values $\chi^{2}/{\mathrm{d.o.f}}$ for the fit functions in the regions of weak~\eq{eq:G:fit:weak} and strong~\eq{eq:G:fit:strong} magnetic field (separated by the double vertical line) vs the magnetic field $eB$ for $14^{4}$ lattice.}
\label{tbl:chi2}
\end{table}

The function~\eq{eq:G:fit:weak} which works surprisingly well at weak values of the magnetic field~\eq{eq:B:quant} with $k=0,1,2$. The values of $\chi^2/d.o.f.$ are shown in Table~\ref{tbl:chi2}. The heuristic function fits nicely our numerical data for the $G_+$ correlator~\eq{eq:G:pm} in the weak field domain while at larger fields ($B \geqslant 1.06 \, {\mathrm{GeV}}^{2}$) the fit gives unacceptably high values of $\chi^2/d.o.f.$ The best fit values for the parameter $\gamma$ are $\gamma \sim 5 \dots 7$. The $k=0,1$ (k=2) fits exclude the short distance separations $z \leqslant a$ ($z \leqslant 2 a$) and their periodic mirrors. The data and the best fits are shown in Fig.~\ref{fig:weak}.

\begin{figure}[!thb]
\begin{center}
\includegraphics[scale=0.9,clip=false]{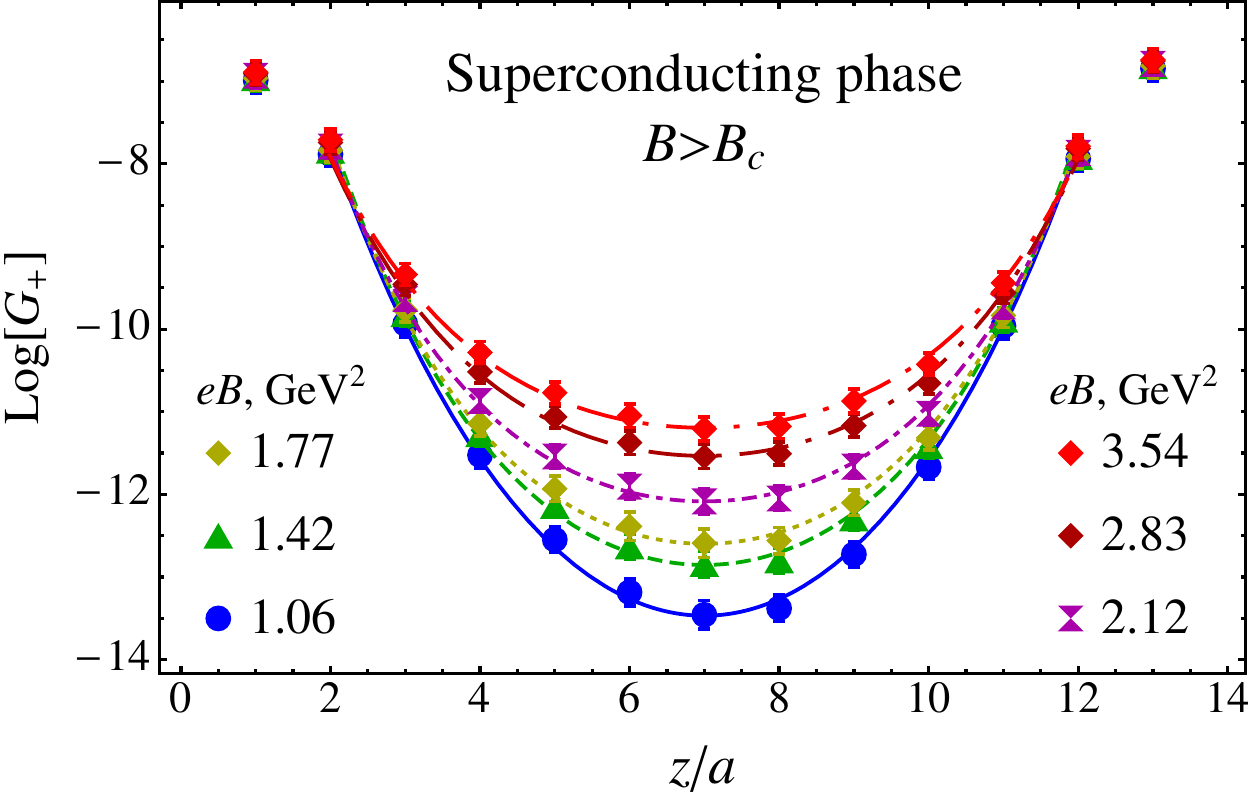}
\end{center}
\caption{The $G_+$ correlator~\eq{eq:G:pm} in the superconducting (strong magnetic field) phase. The lines are the best fits by Eq.~\eq{eq:G:fit:strong}.}
\label{fig:strong}
\end{figure}

We have found that in the high-strength region, $e B > 1 \, {\mathrm{GeV}}^2$, the numerical data for the correlator $G_+(z)$ can well be described by another heuristic function
\beqn
& & G^{\mathrm{fit, strong}}_+ (z) = \eta^2 e^{ - V(z)}\,, 
\label{eq:G:fit:strong} \\
& & V(z) = C_{\mathrm{strong}} \,  e^{-\mu L/2} \cosh [\mu (|z| - L/2)] \,,
\label{eq:V}
\eeqn
for all separations $z$ excluding the ultraviolet region with $z \leqslant a$ (and its periodic mirror). In Eqs.~\eq{eq:G:fit:strong} and \eq{eq:V} $\mu > 0$, $\eta > 0$ and $C_{\mathrm{high}} < 0 $ are the fitting parameters. The $G_+$ correlator in the strong magnetic field region and the corresponding best fits~\eq{eq:G:fit:strong} are shown in Fig.~\ref{fig:strong}. 

\begin{figure}[!htb]
\begin{center}
\includegraphics[scale=0.9,clip=false]{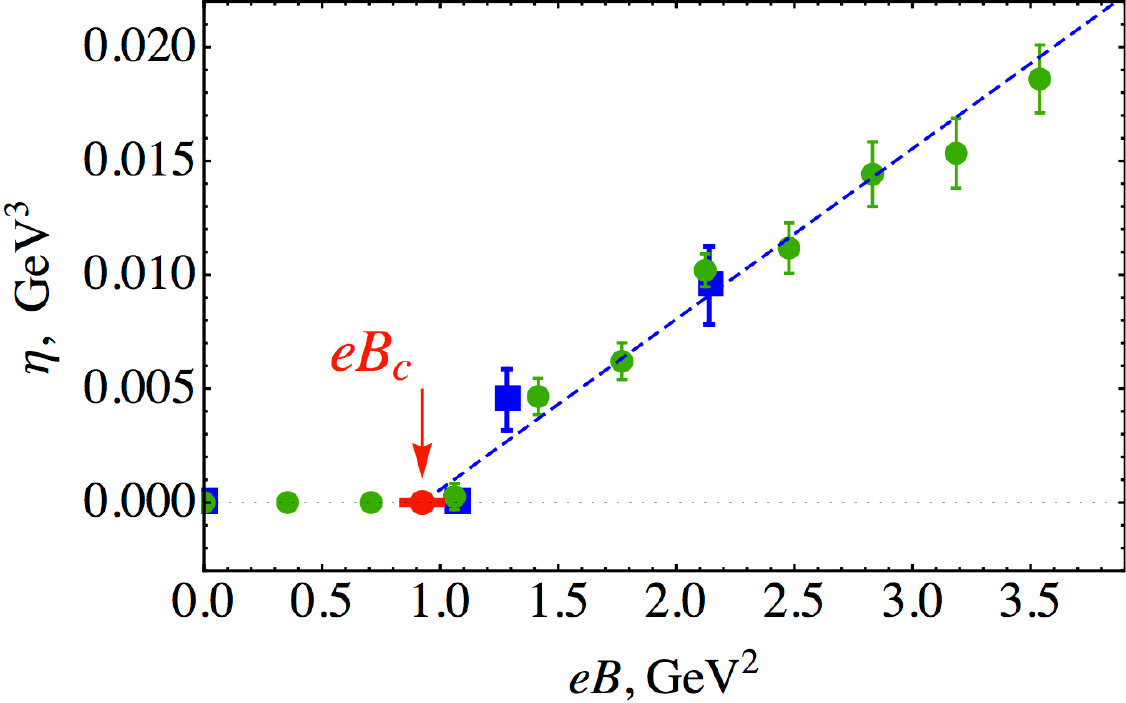}
\end{center}
\caption{The superconducting condensate $\eta = |\langle \rho \rangle|$ of the charged $\rho$ mesons as the function of the magnetic field $B$. The green points correspond to the condensate calculated for small lattice $14^{4}$, while the blue squares represent the data extrapolated to an infinite volume $L \to \infty$. The dashed blue line is the fit by the linear function~\eq{eq:eta:linear}. The red arrow marks the point of the insulator--superconductor phase transition~\eq{eq:Bc}.}
\label{fig:condensate}
\end{figure}

The parameter $\eta$ in the fitting function~\eq{eq:G:fit:strong} plays a role of the charged $\rho$-meson condensate, 
$\eta \equiv |\langle \rho \rangle|$, because in the thermodynamic limit, $L \to \infty$, the function $V(z)$ reduces to an exponential $e^{- \mu |z|}$ so that  
\beqn
 \lim\limits_{z\to \infty} \lim\limits_{l_s \to \infty}  G^{\mathrm{fit, strong}}_+ (z) = \eta^2\,.
\eeqn
Thus, the fits of the $G_+$ correlators provide us with the values of the condensate of the charged $\rho$ mesons, Fig.~\ref{fig:condensate}. The corresponding values of $\chi^2/d.o.f.$ are shown in Table~\ref{tbl:chi2}. In the weak field domain the fitting does not converge properly due to presence of flat directions in the fitting parameter space.

In the weak field region the $\rho$ meson condensate $\eta = |\langle \rho \rangle|$ vanishes, while at higher values of the magnetic field the condensate deviates spontaneously from zero signaling the presence of the superconducting phase. We find that the dependence of the $\rho$--meson condensate on the magnetic field can be described by the linear function:
\beqn
\eta (B) = C_\rho \cdot (eB - eB_c)\,, \qquad B \geqslant B_c\,.
\label{eq:eta:linear}
\eeqn
The best linear fit (shown by the dashed line in Fig.~\ref{fig:condensate}) of the condensate allows us to determine the critical magnetic field of the insulator-superconductor transition,
\beqn
e B_c = 0.924(77)\, {\mbox{GeV}}^2\,,
\label{eq:Bc}
\eeqn
or $B_c = (1.56 \pm 0.13)\cdot 10^{16}\, {\mbox{T}}$, in satisfactory agreement with the theoretical relation~\eq{eq:Bc:theor}, $e B_c = m^2_\rho$, for the quenched mass of the $\rho$ meson
in $SU(2)$ lattice gauge theory, $m_{\rho} \sim 1.1 \,{\mathrm{GeV}}^{2}$ \cite{Stewart:1998hk}. The prefactor in Eq.~\eq{eq:eta:linear} is $C_\rho = 7.5(5) \, {\mathrm{MeV}}$. 
Notice that in our quenched model the exponent in Eq.~\eq{eq:eta:linear} is $\nu=1$ while the mean field methods both in the bosonic VMD model~\cite{ref:PRD} and in the fermionic NJL model~\cite{ref:PRL} predict $\nu=1/2$ (so that theoretically $\eta \sim \sqrt{B - B_c}$ for $B \geqslant B_c$).

\begin{figure}[!thb]
\begin{center}
\includegraphics[scale=0.8,clip=false]{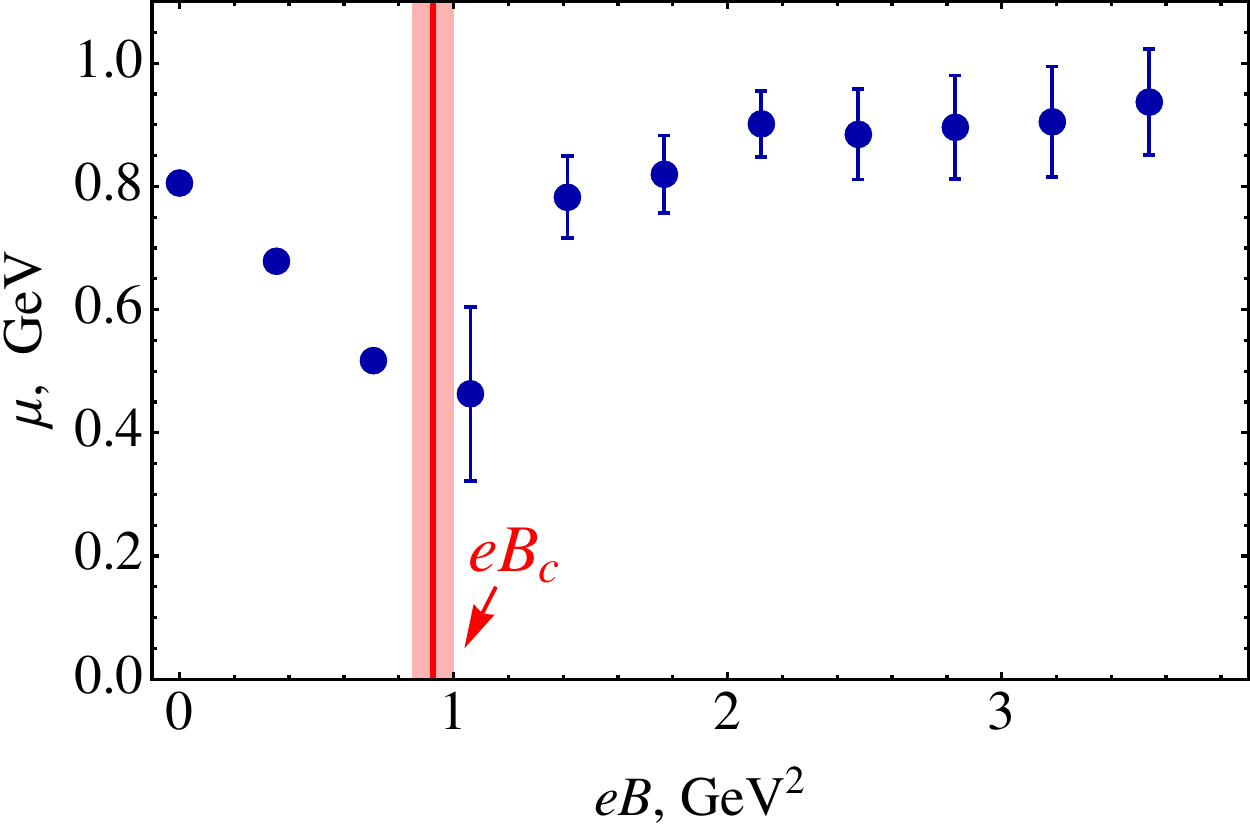}
\end{center}
\caption{The massive parameter $\mu$ corresponding to the best fits \eq{eq:G:fit:weak} and \eq{eq:G:fit:strong}, \eq{eq:V}.}
\label{fig:mu}
\end{figure}

The behaviour of the fitting parameter $\mu$ in the fitting functions at both sides of the critical phase transition~\eq{eq:G:fit:weak} and \eq{eq:G:fit:strong}, \eq{eq:V}, are shown in Fig.~\ref{fig:mu}.

The unusual forms of the fit functions \eq{eq:G:fit:weak} and \eq{eq:G:fit:strong} is used to absorb the finite volume effects. In order to support our small-volume results we have performed an infinite--volume extrapolation of the condensate $\eta$ obtained at larger lattices ($l = 17 \dots 21$ with $L \approx 1.65 \dots 2.2\,\mathrm{fm}$). The condensate was obtained by fitting of the numerical data by the standard function,
\beqn
G^{\mathrm{fit}}_+(z,L) = A  \cosh [m (z - l_{s}/2)] + \eta^{2}(L)\,.
\label{eq:G:extr}
\eeqn
where $A$, $m$ and $\eta$ are the fitting parameters. The fitting parameters and parameters of the lattice at $B \neq 0$  are shown in Table~\ref{tbl:Ls}.

\begin{table}
\begin{tabular}{c | c c c c | c c | c }
$eB$, GeV${}^2$ & $L$, fm & $a$, fm & $l_s$ & $k$ & $\eta$, GeV${}^3$ & $m$, GeV & $\chi^2/$d.o.f.\\
\hline
1.07	 &	1.654 &	0.0973  &	17 &	4 &	0.00626(5)	 &	3.1(1) &	0.59	\\
1.07	 &	1.849 &	0.1027 &	18 &	5 &	0.00501(4) &		3.1(2) &	0.48	\\	
1.07	 &	2.027 &	0.1126 &	18 &	6 &	0.00392(4) &		2.5(3) &	0.56	\\
1.07	 &	2.189 &	0.1152 &	19 &	7 &	0.00353(4) &	 	2.5(2) &	0.87	\\
\hline
1.28	 &	1.688 &	0.0993 &	17 &	5 &	0.00753(4)	 &	3.4(1) &	0.22	\\
1.28	 &	1.849 &	0.1027 &	18 &	6 &	0.00585(4)	 &	3.0(1) &	0.41	\\
1.28	 &	1.998 &	0.111 &	18 &	7 &	0.00446(5)	 &	2.6(2) &	0.49	\\
1.28	 &	2.135 &	0.1186 &	18 &	8 &	0.00486(5)	 &	2.2(3) &	0.33	\\
1.28	 &	2.136 &	0.1124 &	19 &	8 &	0.00454(5)	 &	2.3(2) &	0.61	\\
\hline
2.14	 &	1.754 &	0.1032 &	17 &	9 &	0.01257(7)	 &	2.5(1) &	0.20	\\
2.14	 &	1.849 &  0.1027 &	18 &	10 &	0.01028(6)	 &	2.5(2) &	0.33	\\
2.14	 &	1.940 &	0.1078 &	18 &	11 &	0.00981(8)	 &	2.2(2) &	0.41	\\
2.14	 &	2.025 &	0.1066 &	19 &	12 &	0.00881(7)	 &	2.2(2) &	0.54	\\
2.14	 &	2.109   &	0.111 &	19 &	13 &	0.01001(6)	 &	2.1(3) &	0.32	
\end{tabular}
\caption{The parameters of the lattices used in the thermodynamic extrapolation for nonzero values of $B$ and the corresponding best fit parameters obtained with the help of Eq.~\eq{eq:G:extr}. The data is visualised in Fig.~\ref{fig:extr}.}
\label{tbl:Ls}
\end{table}

\begin{figure}[!thb]
\begin{center}
\includegraphics[scale=0.9,clip=false]{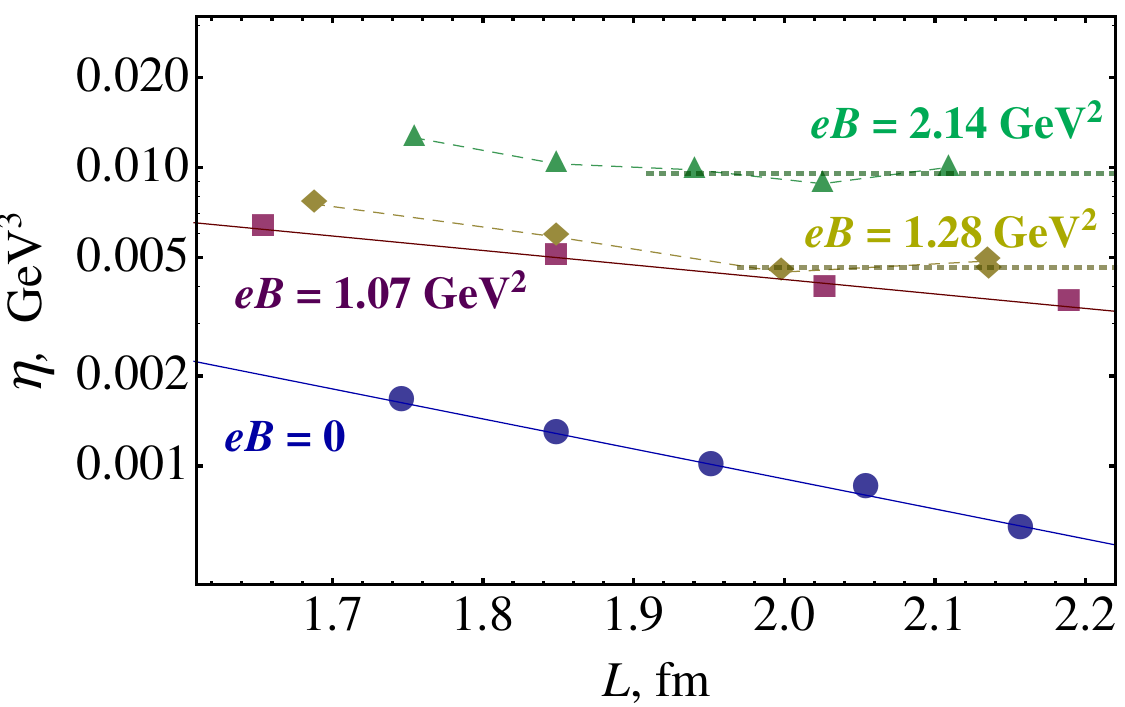}
\end{center}
\caption{The superconducting condensate $\eta(L)$ vs. the lattice size $L$ at fixed values of the magnetic field $B$. The dashed lines are shown to guide eye. The solid lines are the best fits of the data by the exponential function~\eq{eq:exp}.}
\label{fig:extr}
\end{figure}

It turns out that in the insulator phase, $B < B_{c}$, the data for $\eta(L)$ can very well be fitted by the exponentially decaying (in the $L\to\infty$ limit) function
\beqn
\eta^{\mathrm{fit}}(L) = C e^{- L/L_0}\,,
\label{eq:exp}
\eeqn
where $L_0$ and $C$ are fitting parameters. The condensate in the infinite volume tends zero at $B < B_{c}$, as expected. The fits are shown in Fig.~\ref{fig:extr} by the solid lines. The corresponding slopes are $L_0 = 0.42(2)$\,fm and $L_0 = 0.90(8)$\,fm for $eB = 0$ and $eB = 1.07$\,GeV${}^2$, respectively.

At higher values of $B$, the condensate shows plateaux as $L$ increases. We get the $L \to \infty$ extrapolation for the condensate by averaging the data for $\eta(L)$ at two largest values of the lattice size $L$ (the horizontal dotted lines in Fig.~\ref{fig:extr}). The extrapolated data -- shown by the blue squares in Figure~\ref{fig:condensate} -- agrees quantitatively well with our small-volume analysis. The nonzero values of the extrapolated condensates are $\eta = 0.0046(4)$\,GeV${}^2$ and $\eta = 0.0095(1)$\,GeV${}^2$ for $eB = 1.28$\,GeV${}^2$ and $eB = 2.14$\,GeV${}^2$, respectively.

Thus, our numerical results support the theoretical prediction of Refs.~\cite{ref:PRD,ref:PRL} that the superconducting charged condensate of the $\rho$ mesons forms spontaneously at strong magnetic field. Using the simulations of the quenched QCD vacuum, we determined the critical magnetic field~\eq{eq:Bc} which is remarkably close to the theoretical prediction~\eq{eq:Bc:theor}.  

Finally, we would like to stress that theoretical calculations show that the condensate in the vacuum ground state should be an inhomogeneous function of spatial coordinates~\cite{ref:PRD,ref:PRL}. The ground state can be represented as a coherent static lattice of the topological (vortex-like) defects in the $\rho$--meson condensates, the $\rho$ vortices, which are directed along the magnetic field axis. Qualitatively, the $\rho$--vortex state is very similar to the Abrikosov vortex lattices observed in the type--II superconductors in a background of a strong magnetic field~\cite{Abrikosov:1956sx}.

It turns out, however, that in the QCD vacuum the energy gap between the lowest vortex energy state (given by a triangular vortex lattice) and excited vortex lattice states is parametrically very small~\cite{Chernodub:2011gs}, implying that the spatial lattice order of the vortex state may be destroyed by quantum (or thermal) fluctuations. The latter fact indicates that the actual vortex structures in the superconducting phase may resemble a much less ordered but persistent ``spaghetti state'', where the correlation functions, given by Eq.~\eq{eq:G:pm} and/or Eq.~\eq{four_fermion_vev},  get additional suppression factors due to almost random vortex motion. The investigation of the detailed features of the superconducting ground state is currently underway~\cite{ref:preparation}.

\vskip 3mm 

The authors are obliged to A.S. Gorsky for interesting discussions. The work we supported by Grant "Leading Scientific Schools" No. NSh-6260.2010.2, RFBR-11-02-01227-a, Federal Special-Purpose Program "Cadres" of the Russian Ministry of Science and Education, by a grant from FRRC, and by grant No. ANR-10-JCJC-0408 HYPERMAG (France). Numerical calculations were performed at the  ITEP system Stakan (authors are much obliged to A.V. Barylov, A.A.  Golubev, V.A. Kolosov, I.E. Korolko, M.M. Sokolov for help), the MVS 100K at Moscow Joint Supercomputer Center and at Supercomputing Center of the Moscow State University.

\bibliographystyle{model1a-num-names}

\end{document}